\documentstyle[12pt]{article}
\newcommand{\preprint}[1]{\begin{table}[t]        
            \begin{flushright}                    
            \begin{large}{#1}\end{large}          
            \end{flushright}                      
            \end{table}}                          

\begin{document}
\preprint{LA-UR-98-2878}
\title{\it Electromagnetic duality and light-front coordinates. }
\author{ Martina M. Brisudov\' a \\
{\it Theoretical Division, T-5}\\ {\it Los Alamos National Laboratory, Los 
Alamos, NM 87 545} }
\date{\today}
\maketitle
\abstract{We review the light-front Hamiltonian approach for the Abelian gauge 
theory in 3+1 
dimensions, and then study electromagnetic duality in this framework. }
\newpage

\section{Introduction.}
In recent years we have witnessed a resurrection of interest in light-front 
Hamiltonian  physics 
in two areas. The first one, the new nonperturbative 
approach to QCD \cite{thelongpaper}, is related 
to the original application of light-front coordinates \cite{ancient}, i.e.
 hadron spectroscopy, 
and the other comes  
 from string theory \cite{thorn}. 
The popular M-theory  \cite{Mtheory} is 
formulated in light-front coordinates \cite{motl}. 
With the rise of the second application, some rather academic questions 
became of interest.For example, dualities in string theories are one of
 the most powerful tools, yet not much is known about how they work on the 
light front. In this paper, we attempt to study one of the simplest cases of
 known dualities - 
the electromagnetic duality in the Abelian gauge field theory in 3+1 dimensions. 

 Susskind has conjectured \cite{iowa} that since light-front
 coordinates are non-local in the longitudinal direction, 
it might be possible to formulate a light-front theory
 with both electric and magnetic sources without having to introduce 
any additional non-localities corresponding to Dirac
 strings \cite{goddardolive}. He observed that the role of
 electric and magnetic fields reverses in the light-front
 Hamiltonian (which contains only the physical, transverse fields) when
 the original fields are replaced by (transverse) fields 
perpendicular to them. He concluded that the above described
 transformation of fields is the electromagnetic duality on the light front. 
Then he suggested that magnetic sources be added into the Hamiltonian 
by symmetry.

In this paper we investigate this idea. The paper is 
organized as follows: Since there are some misconceptions in the 
literature, and since each beginning researcher in this field 
has to set up his/her own notes 
on the light-front conventions, we 
summarize in the section 2 formalisms of light-front 
coordinates, free Abelian fields, and we establish the 
connection between the components of the $F^{\mu \nu} $ tensor 
and electric and magnetic fields. We show how  classical electric
 sources can be added to the theory.
For completeness, we list the surface terms  even though 
they do not enter the calculation 
presented here, and  
we list some manipulations with the $(\partial^+)^{-1}$ operator. Further, we 
wish to mention that 
there are, in general,  problems regarding other than $+$-components of 
light-front currents,  even though this does not affect our calculation 
since we restrict ourselves to external classical currents. The section 2
is rather formal; a reader familiar with the light front may want 
to skip most of it. Section 3 is 
 devoted to Susskind's idea. The last section contains our conclusions.

\section{Light-front field theory: Formalities}

In a light-front quantum field theory, fields are quantized at an equal
 light-front time \cite{dirac}. Advantages of a light-front formulation are:
 The light-front has the largest kinematic subgroup of Lorentz generators, 
boosts are kinematic \cite{coester}, 
and the light-front vacuum can be decoupled from the 
physical states by imposing a longitudinal momentum 
cutoff \cite{thelongpaper}. The price to pay is more 
complicated renormalization, rotations involving the z-axis are dynamical, and 
as a consequence the physical picture is less intuitive for nonrelativistic 
systems at rest that are naturally described in equal-time coordinates.
On the other hand, it is a natural framework for highly relativistic
 systems (e.g. description of deep inelastic scattering).

\subsection{ Light-front coordinates}
 Dirac showed that it is possible to formulate relativistic 
dynamics in coordinates other than the usual equal-time form in which everything 
is 
expressed in terms of  dynamical 
variables at one instant of time (hence {\it instant form}). The other 
forms he found are the {\it point form} and the {\it front form} \cite{dirac}.

In the front form (or light-front) 
$x^+ = t + z$ plays the role of time. The remaining coordinates,
 $x^- = t-z$ and 
$x^{\perp} \equiv (x^1, x^2)$, are spatial. Given a four-vector $a$, its 
components in light-front coordinates are: 
\footnote{There are two slightly different conventions regarding the $+,-$ 
components. 
The other one  differs from the one used here by a factor of 
$\sqrt{2}$: $a^{\pm} = (a^0 \pm a^3)/\sqrt{2}$, so that the metric tensor 
$g^{+-}=1$. It is, therefore, a good idea to check the definitions of 
coordinates before comparing any results. }
\begin{eqnarray}
a^- & = & a^0 - a^3 , \nonumber\\
a^+ & = & a^0 + a^3 , \nonumber\\
a^{\perp} & = &(a^1, a^2) .
\end{eqnarray}
A scalar product of two four-vectors $a,b$  is:
\begin{eqnarray}
 a_{\mu} b^{\mu}
  = {1\over{2}} a^+ b^- + {1\over{2}} a^- b^+ - a^{\perp}\cdot b^{\perp}.
\end{eqnarray}

The metric tensor in light-front coordinates is:
\begin{eqnarray}
g_{\mu \nu} = \left( \begin{array}{cccc}
 g_{++} & g_{+-} & g_{+1} & g_{+2} \\
g_{-+} & g_{--} & g_{-1} & g_{-2} \\
g_{1+} & g_{1-} & g_{11} & g_{12} \\
g_{2+} & g_{2-} & g_{21} & g_{22} \end{array} 
\right) 
=\left( \begin{array}{cccc}0 & 1\over{2} & 0 & 0\\
1\over{2} & 0 & 0 & 0 \\
0 & 0 & -1 & 0 \\
0 & 0 & 0 &-1 \end{array} \right)
\end{eqnarray}
and
\begin{eqnarray}
g^{\mu \nu} = \left( \begin{array}{cccc}
 g^{++} & g^{+-} & g^{+1} & g^{+2} \\
g^{-+} & g^{--} & g^{-1} & g^{-2} \\
g^{1+} & g^{1-} & g^{11} & g^{12} \\
g^{2+} & g^{2-} & g^{21} & g^{22} \end{array}
\right) 
=\left( \begin{array}{cccc} 0 & 2 & 0 & 0\\
2 & 0 & 0 & 0 \\
0 & 0 & -1 & 0 \\
0 & 0 & 0 &-1 \end{array}\right)
\end{eqnarray}

We can now write down the derivatives with respect to coordinates:
\begin{eqnarray}
\partial ^- & = & {\partial \over{\partial x_-}}
 = 2 {\partial \over{\partial x^+}} = \partial ^0 - \partial ^3 , \nonumber\\
\partial ^+ & = &{\partial \over{\partial x_+}}
 = 2 {\partial \over{\partial x^-}} =
 \partial ^0 + \partial ^3 , \nonumber\\
\partial ^{\perp} & = & (\partial ^1, \partial ^2) .
\end{eqnarray}
 $\partial ^-$ is the time derivative, the remaining derivatives are 
spatial.

The four-dimensional volume element is:
\begin{eqnarray}
\big[ d^4x \big] = {1\over{2}} dx^+dx^-d^2x^{\perp} .
\end{eqnarray}

Let $p= (p^-, p^+, p^1, p^2)$ be the
 four-momentum of a  free particle with mass $m$ in light-front coordinates. 
Then
\begin{eqnarray}
p_{\mu} x^{\mu} 
  = {1\over{2}} p^+ x^- + {1\over{2}} p^- x^+ - p^{\perp}\cdot x^{\perp}.
\end{eqnarray}
 $p^+$ is the {\it longitudinal} momentum, $p^1$ and $p^2$ are 
the {\it transverse} momenta, and $p^-$ is {\it the light-front energy}:
\begin{eqnarray}
p^- ={{p^{\perp}}^2 + m^2 \over{p^+}} .
\end{eqnarray}

The Lorentz invariant momentum integration element is obtained as follows:
\begin{eqnarray}
\big[ d^4q \big] \, 2 \pi \, \delta (m^2 -q^2) = {1\over{2}} 
{dq^- dq^+ d^2q^{\perp}\over{ (2\pi)^3}} \delta(m^2 -q^+ q^- +q^{\perp \  2})
={dq^+ d^2q^{\perp} \over{2 (2 \pi)^3 q^+}}  .
\end{eqnarray}

The light-front energy is well defined apart from peculiar modes which have 
zero longitudinal momentum (so-called {\it zero modes}).
$p^+$, which is equal to $p^0 +p^3$, satisfies 
 $p^+ \geq 0$. This means that in the
 vacuum all particles must have precisely zero 
longitudinal momentum. From the expression for the light-front energy we can 
see that the energy diverges as $p^+ \rightarrow 0$ for massive particles. 
For massless particles, the light-front energy can be finite even at
$p^+ =0$, but the vacuum can be made trivial by imposing a small 
longitudinal momentum cutoff, e.g. requiring that all longitudinal momenta
 satisfy $p_i^+ >\epsilon$. 
 Another frequently used method of regularization is a 
discretized light-cone quantization (DLCQ) \cite{DLCQ} which removes both 
ultraviolet and infrared divergences. The physics of $p^+=0$ cannot be 
recovered by renormalization 
with respect to high energy states, and has to be added by hand  
using  counterterms. The so-called ``constraint zero mode'' \cite{zero} 
is a specific counterterm consequent of DLCQ, and it does
 not require a 
nontrivial vacuum structure.

\subsection{Free Abelian gauge fields}
 Let us start with pure electromagnetism. The Lagrangian density is:
\begin{eqnarray}
{\cal L} = -{1\over{4}} F_{\mu \nu} F^{\mu \nu} ,
\end{eqnarray}
where 
\begin{eqnarray}
F^{\mu \nu} = \partial ^{\mu} A^{\nu} - \partial ^{\nu} A^{\mu}.
\end{eqnarray}
In a light-front formulation, indices $\mu$, $\nu$ run through
 $+, -$, and $ \perp =(1,2)$.

In {\it light-front gauge}, $A^+ =0$, and the Lagrangian density reduces to:
\begin{eqnarray}
{\cal L} = {1\over{8}} (\partial ^+ A^-)^2 
 + {1\over{2}}\partial ^+ A^i \partial ^-A^i -
{1\over{2}} \partial ^+ A^i \partial ^i A^-
- {1\over{4} } \left(\partial ^i A^j -\partial ^j A^i\right)^2 .
\end{eqnarray}
The Lagrangian density does not contain a time-derivative of $A^-$ so it is 
immediately obvious that this component of $A^{\mu}$ is not dynamical.
Indeed, conjugate momenta, $\Pi $, to the fields  are:
\begin{eqnarray}
\Pi _{A^i} = {\partial {\cal L} \over{\partial \ \partial ^- A^i}} & = &
{1\over{2}} \partial ^+ A^i ,\nonumber\\
\Pi _{A^-} = {\partial {\cal L} \over{\partial \ \partial ^- A^-}}
& = & 0 .
\end{eqnarray}
$A^-$ can be eliminated using the equations of motion:
\begin{eqnarray}
{\partial {\cal L} \over{ \partial A^-}} = \partial ^{\mu} 
{\partial {\cal L} \over{ \partial \ \partial ^{\mu} A^-} },
\end{eqnarray}
leading to
\begin{eqnarray}
\left( \partial ^+ \right)^2 A^-  = 2 \partial ^+ \partial ^i A^i .
\end{eqnarray}
Apart from zero modes (i.e. $p^+=0$ states introduced above), 
$\partial ^+$ can be inverted, and $A^-$ is then 
given by:
\begin{eqnarray}
A^- = {2\over{\partial ^+}}  \partial ^i A^i .
\end{eqnarray}

To proceed further, some manipulations with $(\partial^+)^{-1}$ are needed.
Up to a constant in $x^-$ which can depend on remaining coordinates, the 
operator $(\partial^+)^{-1}$ is defined as follows:
\begin{eqnarray*}
{1\over{\partial^+}}f(x^-) \equiv \int {dy^-\over{4}} \epsilon (x^- -y^-)
f(y^-) , 
\end{eqnarray*}
where $\epsilon(x) = 1$, if $x>0$, and $\epsilon(-x) = - \epsilon(x)$, and 
$f(x^-)$ is an arbitrary function.
Using the properties of $\epsilon (x)$ it is straightforward to find:
\begin{eqnarray*}
\int d^3x \left({1\over{\partial^+}}f(x)\right)^2 
= - \int d^3x f(x) \left({1\over{\partial^+}}\right)^2f(x),
\end{eqnarray*}
\begin{eqnarray*}
\int d^3x f(x) \left({1\over{\partial^+}}g(x)\right)
= -\int d^3x  \left({1\over{\partial^+}}f(x)\right)g(x).
\end{eqnarray*}

Substituting for $A^-$, and using the properties of the 
operator $(\partial^+)^{-1}$ shown above,  gives the Lagrangian in terms of 
physical degrees of freedom:
\begin{eqnarray}
{\cal L} = {1\over{2}} \partial ^+ A^i \partial ^- A^i
- {1\over{2}} \left( \partial ^i A^i\right) ^2
-{1\over{4}} \left( \partial ^i A^j -\partial ^j A^i \right)^2
+ {\rm surface \ \ terms} ,
\end{eqnarray}
where the surface terms,
\begin{eqnarray}
-\left\{ \partial ^+ \left( A^j \partial ^j {1\over{\partial ^+}} 
\partial ^i A^i \right) 
- \partial ^j \left( A^j \partial ^i A^i \right) \right\} ,
\end{eqnarray}
are traditionally dropped.
$F^{\mu \nu} $ in terms of $A^i$ is:
\begin{eqnarray}
F^{+-} =  2 \partial ^i A^i , & \  &
F^{+i}  =  \partial ^+ A^i ,  \nonumber\\
F^{-i} = \partial ^- A^i -
 \partial^i {2\over{\partial ^+}}\partial ^j A^j ,& \  & 
F^{ij}=  \partial ^i A^j - \partial ^j A^i .
\end{eqnarray}
Let us also introduce the dual tensor 
$\tilde{F}^{\mu \nu} = 1/2 \epsilon ^{\mu \nu \lambda \rho}  F_{\lambda \rho}$,
where $\epsilon ^{\mu \nu \lambda \rho}$ is totally antisymmetric,
\begin{eqnarray}
\epsilon^{+-12} \equiv 2  ,
\end{eqnarray}
so that it satisfies
\begin{eqnarray}
\epsilon ^{\mu \nu \alpha \beta} \epsilon _{\mu} ^{\  
\nu ' \alpha ' \beta '}
 & = & g^{\nu \alpha '} g^{\alpha \nu '} g^{\beta \beta '}
+ g^{\nu \nu '} g^{\alpha \beta '} g^{\beta \alpha '}
+ g^{\nu \beta ' } g^{\alpha \alpha ' } g^{\beta \nu '} \nonumber\\
  &  & - g^{\nu \nu '} g^{\alpha \alpha '} g^{\beta \beta '}
- g^{\nu \beta '} g^{\alpha \nu ' } g^{\beta \alpha '}
- g^{\nu \alpha '} g^{\alpha \beta '} g^{\beta \nu '} .
\end{eqnarray}
Then,
\begin{eqnarray}
\tilde{F}^{+-} =  \epsilon_{ij} F^{ij}, 
 & \  &
\tilde{F}^{+i}  =  \epsilon_{ij} F^{+j}, 
 \nonumber\\
\tilde{F}^{ij}  =  -{1\over{2}}\epsilon_{ij}F^{+-}, 
 & \  &
\tilde{F}^{-i} = - \epsilon_{ij} F^{-j}.
\end{eqnarray}
where  $\epsilon_{12}=1$ is 
antisymmetric. Let us note that $F$ and 
$\tilde{F}$ are related by electromagnetic 
duality $\vec{B} \rightarrow \vec{E}$,
$\vec{E} \rightarrow -\vec{B}$. The connection between electric and 
magnetic fields $\vec{E}$ and $\vec{B}$ and the tensor $F^{\mu \nu}$ given 
in light-front coordinates is shown in the next section.

\subsection{ Connection between electric and magnetic fields $\vec{E}$ and 
$\vec{B}$ and the tensor $F^{\mu \nu}$ in light-front coordinates.}

The connection between $\vec{B}$, $\vec{E}$ and $F^{\mu \nu}$ 
can be established using the definition of the potential:
\begin{eqnarray}
\vec{E} & = &-\vec{\nabla} A^0 - {\partial \over{\partial t}}\vec{A} , 
\nonumber\\
\vec{B} & = & \vec{\nabla} \times \vec{A}  . 
\end{eqnarray}
Substituting:
\begin{eqnarray}
A^0  =  {1\over{2}} \left( A^+ +A^-\right),  & \  &
A^3  =  {1\over{2}} \left( A^+ -  A^- \right), \nonumber\\
{\partial \over{\partial t}} = 
\partial ^0 =  {1\over{2}} \left( \partial ^+ + \partial ^-\right), 
 & \  &
{\partial \over{\partial z}} = - \partial ^3
  =  - {1\over{2}} \left( \partial ^+ - \partial ^- \right),
\end{eqnarray}
we obtain:
\begin{eqnarray}
E^i  =  -{1\over{2}}\left( F^{+i} +F^{-i} \right) ,
& \  &
E^z =  {1\over{2}} F^{+-}, \nonumber\\
B^i  =  \epsilon_{ij} {1\over{2}} \left( F^{+j} - F^{-j} \right),
& \  &
B^z  =  -{1\over{2}} \epsilon _{ij} F^{ij},
\end{eqnarray}
or, 
\begin{eqnarray}
F^{+-}  =  2E^z , & \  &  \tilde{F}^{+-}  =  -2B^z \nonumber\\
F^{+i}  =  -\left( E^i + \epsilon _{ij} B^j \right), & \  & 
\tilde{F}^{+i}  =  \left( B^i - \epsilon _{ij} E^j \right) ,\nonumber\\
F^{ij}  =  -\epsilon _{ij} B^z, 
& \  & \tilde{F}^{ij}  =  -\epsilon _{ij} E^z, 
\nonumber\\
F^{-i} =  -\left( E^i - \epsilon _{ij} B^j \right) , & \  & 
\tilde{F}^{-i} =  \left( B^i + \epsilon _{ij} E^j \right) ,
\end{eqnarray}
where $i,j$ are transverse indices ($i,j =(1,2)$).
 These definitions ensure that the Lagrangian 
equations of motion give the correct set of Maxwell's equations. In 
ref. \cite{xy} the magnetic and electric fields are defined differently, in 
particular, in analogy with  equal time, $E^{\mu} = 1/2 F^{+\mu}$ and 
$B^- =F^{12}$, but  this is
misleading, because these definitions do not lead to Maxwell's equations. 
Moreover, $\vec{E}$ and $\vec{B}$ are not four-vectors, $E^0$ and $B^0$ are 
not defined, so there is no natural way  to form 
the minus and plus components.

\subsection{ Adding classical electric sources}
In this section we add classical electric sources.
The Lagrangian density in the presence of classical sources $j_{\mu}$ is:
\begin{eqnarray} 
{\cal L}  = -{1\over{4}} F^{\mu \nu} F_{\mu \nu} - j_{\mu} A^{\mu}
\end{eqnarray}
As before, $A^- $ is not dynamical and it can be eliminated using the 
equations of motion, leading to:
\begin{eqnarray}
A^- & = & {2\over{\partial ^+}} \left( \partial ^i A^i - {1\over{\partial ^+}}
j^+ \right)  \nonumber\\
 & = & A^-_{\rm free} - {2\over{(\partial ^+)^2}}j^+  .
\end{eqnarray}
Replacing $A^-$ modifies $F^{\mu \nu}$, in particular, the $+$-component of the 
current is absorbed into $F^{+-}$ and $F^{-i} $:
\begin{eqnarray}
F^{+-} & = & F_{\rm free}^{+-} -{2\over{\partial ^+}} j^+ ,  \nonumber\\
F^{-i} & = & F_{\rm free}^{-i} -\partial ^i {2\over{(\partial ^+)^2}}j^+  , 
\end{eqnarray}
where $F_{\rm free}^{\mu \nu}$ is given in previous sections. The remaining two 
components of $F^{\mu \nu}$  are unchanged. The Lagrangian density then reads:
\begin{eqnarray}
{\cal L} = {1\over{2}} \partial ^+ A^i \partial ^- A^i
- {1\over{2}} \left( \partial ^i A^i -{1\over{\partial ^+}} j^+  \right) ^2
-{1\over{4}} \left( \partial ^i A^j -\partial ^j A^i \right)^2
+j^{\perp} A^{\perp} \nonumber\\
+{\rm surface \ \ terms} ,
\end{eqnarray}
and the surface terms are modified also:
\begin{eqnarray}
\lefteqn{{\rm surface \ terms} =}\nonumber\\
& - & \partial ^+ \left[ A^j \partial ^j {1\over{\partial ^+}} 
\left( \partial ^i A^i - {1\over{\partial^+}}j^+\right) \right]
- \partial ^j \left[ A^j \left( \partial ^i A^i -{1\over{\partial^+}}
\right)\right] \nonumber\\
& + & {1\over{\partial^+}}\left[ j^+\left( \partial^i A^i -{1\over{\partial^+}}
j^+ \right)\right]  . 
\end{eqnarray}

The Lagrangian  equations of motion are:
\begin{eqnarray}
\partial _{\mu} F^{\mu i} = j^i ,
\end{eqnarray}
$\partial _{\mu}F^{\mu +} =j^+$ is satisfied identically, and using equations of 
motion 
it can be shown that 
\begin{eqnarray*}
\partial _{\mu}F^{\mu -} = -{2\over{\partial ^+}}\left[ {1\over{2}}
\partial ^- j^+ - \partial ^i j^i \right]  ,
\end{eqnarray*}
which implies a continuity equation for $j^\mu$.

The Hamiltonian density in the presence of classical sources is:
\begin{eqnarray}
{\cal H} ={1\over{2} } (\partial ^i A^i-{1\over{\partial ^+}} j^+  )^2 +
{1\over{4}}(\partial^i A^j -\partial^j A^i)^2 -j^{\perp} A^{\perp}   ,
\end{eqnarray}
and the fields $A^i$ can be quantized as if they were free.

\section{Electromagnetic duality}

In this section we investigate the question of whether 
it is possible to formulate electromagnetic duality as a 
transformation of the potential $A^{\perp}$ itself rather than 
the field strength tensor and its dual. Given the Hamiltonian in the 
transverse degrees of freedom, a natural starting point is 
a transformation 
\begin{eqnarray}
A^j \  \  \rightarrow \  \  \tilde{A}^i \equiv -\epsilon _{ij} A^j.
\end{eqnarray}

Indeed, under this transformation  
the first and second term in the free Hamiltonian ``interchange'':
\begin{eqnarray}
{\cal H} =
{1\over{4}}(\partial^i \tilde{A}^j -\partial^j \tilde{A}^i)^2 +
{1\over{2} } (\partial ^i \tilde{A}^i )^2  .
\end{eqnarray}
By comparison with the Hamiltonian including electric sources (see eqn. $(33)$
), 
 it appears that one can by symmetry add 
magnetic sources, as well as the electric sources. In complete analogy  one 
could then expect that the $+$-component of the magnetic current 
$\tilde{j}^{\mu}$ was absorbed into the definition of the field strength tensor 
and/or the dual tensor \cite{iowa}. Taking advantage of the kinematical
 boost invariance (for a review see \cite{coester}), it would be sufficient
to
consider the simple case  of a  magnetic current with  only  the $+$-component
 (the so-called {\it good component}) being non-zero, viz.
\begin{eqnarray*}
{\cal H} = {1\over{2} } (\partial ^i A^i-{1\over{\partial ^+}} j^+  )^2 
+{1\over{2} } (\partial ^i \tilde{A}^i-{1\over{\partial ^+}} \tilde{j}^+  )^2
 -j^{\perp} A^{\perp}. 
\end{eqnarray*}
\noindent  
It is straightforward to show that 
if one proceeds as described, the Hamiltonian leads to the desired equations 
of motion, including the continuity equation for the $-$-component of 
$\tilde{j}$. \footnote{It is not really a mystery - due to the definition 
of the dual tensor as 
$\tilde{F}^{\mu \nu} \equiv 1/2 \epsilon ^{\mu \nu \lambda \rho} F_{\lambda 
\rho}$, it 
follows that $\partial _{\mu} \tilde{F}^{\mu \nu} \equiv 0$. However, 
absorbing the $\tilde{j}^+$ appropriately into the definition 
of $\tilde{F}$ produces a non-zero right-hand side. It is somewhat reminiscent 
of introducing a Dirac string. Note that in our case this 
trick does not work for $\tilde{j}^i \neq 0$.}

The catch is that the Hamiltonian itself is {\it  not} equivalent 
to the complete set of Maxwell equations. It is, rather, the 
Hamiltonian {\it and} the gauge conditions \cite{jacksonBaby}. 
In particular, {\it only }with the gauge conditions $A^+=0$ 
and $A^-= 2 (\partial^+)^{-1} \partial ^i A^i$  are all components of 
the field strength tensor  defined unambiguously.

Let us look at what happens with the field strength tensor and its dual 
under the transformation $(34)$. In order for the transformation $(34)$ 
to be the operation of electromagnetic duality, it has to lead to 
\begin{eqnarray}
-\tilde{F}^{\mu \nu}(A^{\perp}) = F^{\mu \nu} (\tilde{A}^{\perp}),
 \nonumber\\
{F}^{\mu \nu}(A^{\perp}) =\tilde{F}^{\mu \nu} (\tilde{A}^{\perp}).
\end{eqnarray}

However,
\begin{eqnarray}
-\tilde{F}^{+-} & = &  2 \partial^i \tilde{A}^i , \nonumber
\\
-\tilde{F}^{+i} & = &  \partial ^+ \tilde{A}^i ,
 \nonumber\\
-\tilde{F}^{-i} & = & 
\partial ^- \tilde{A}^i
-\partial^i \left( {2\over{\partial ^+}}\partial ^k \tilde{A}^k
\right) - {2\over{\partial ^+}}\Box \tilde{A}^i  , \nonumber\\
-\tilde{F}^{ij} & = & 
= \partial^i \tilde{A}^j - \partial^j \tilde{A}^i 
\end{eqnarray}
 shows that the transformation $(34)$ 
is not quite electromagnetic duality: It works for all components
except  $F^{-i}$. The $\tilde{F}^{-i}$ contains an additional 
term $- 2({\partial ^+})^{-1}\Box \tilde{A}^i $.

 For free fields, the additional term is zero,
 and $(34)$ is therefore electromagnetic duality. Is it possible to 
remove the additional term in general,  realizing the electromagnetic duality
 as a generalization of the original Susskind's suggestion, 
e.g. $(34)$ plus a gauge transformation? 

After fixing the gauge, there is still a residual gauge freedom. 
In order not to disturb the gauge conditions used to derive the Hamiltonian,
 the residual gauge function $\Lambda$ 
has to satisfy
\begin{eqnarray}
\partial ^+ \Lambda  = 0,  & \  &
\partial^- \Lambda  =  {2\over{\partial^+}}(\partial^i)^2 \Lambda .
\end{eqnarray}
Ignoring for a moment the question of zero modes, this implies that 
\begin{eqnarray*}
{2\over{\partial ^+}}\Box \Lambda =0
\end{eqnarray*} 
and thus cannot cancel the unwanted term 
$- 2({\partial ^+})^{-1}\Box \tilde{A}^i $.

We now return to the question of zero modes.  
Since they correspond to a constant in $x^-$, they cannot cancel the 
$- 2({\partial ^+})^{-1}\Box \tilde{A}^i $ term which, 
in general, does depend on $x^-$.

\section{Conclusion and summary}
We reviewed the formalism of Abelian gauge theory in light-front coordinates. We 
argued that while the potential $A^{\mu}$ can be described in light-front 
coordinates, there is no light-front analogue to electric and magnetic fields 
$\vec{E}$, $\vec{B}$ in the sense that if one defines electric and magnetic 
fields as components of the light-front field strenght tensor, the definitions 
do not lead to Maxwell equations, and electromagnetic duality is not realized as 
$\vec{B} \rightarrow \vec{E}$,
$\vec{E} \rightarrow -\vec{B}$. 
 
We then studied electromagnetic duality on the level of fields $A^{\mu}$ (in 
light-front gauge $A^+ =0$). Our study was motivated by the fact that the 
light-front Hamiltonian is in this case expressed in terms of transverse fields 
only, and that by under a specific transformation of transverse fields the 
electric and magnetic terms in the Hamiltonian interchange.

However, the electromagnetic duality in light-front coordinates cannot be 
realized
 by a transformation of transverse fields only. Neither can it be written
 as a transformation of transverse fields plus a gauge transformation, 
not even when the gauge transformation has a zero mode. 
Altering $A^-$ in addition to $A^{\perp}$ is not likely to fix the problem 
either, because it is only one of the two components of the dual tensor 
involving $A^-$ (i.e. $\tilde{F}^{-i}$) that does not transform as desired; 
fixing $\tilde{F}^{-i}$ would spoil the transformation of $\tilde{F}^{-+}$.

To include 
magnetic monopoles one would have to allow for additional non-localities (in 
the gauge function), most likely equivalent to Dirac strings in an equal-time
 theory \cite{goddardolive}.

\section{ Acknowledgments }
My work has been supported by the United States Department of Energy.
I would like to acknowledge L. Susskind for bringing this problem into my 
attention. I am grateful to G. t'Hooft for useful discussions during the 
{\it NATO ASI Workshop on Confinement, Duality, and Non-Perturbative Aspects of 
QCD } held June 23 - July 4, 1997, and to the organizers of the workshop, 
particularly P. van Baal for providing such a stimulating research environment. 
I am also grateful to R. Furnstahl, T. Goldman and R. Perry for 
reading the manuscript. Last but not least, I want to thank M\'{a}ria 
Barn\'{a}\v{s}ov\'{a} for making this work possible.

\end{document}